# Neutron-induced reactions in nuclear astrophysics

René Reifarth[1,*], David Brown[2], Saed Dababneh[3],

Yuri A. Litvinov[4], Shea M. Mosby[5]

[1]*Goethe University Frankfurt, Frankfurt, Germany*

[2]*Brookhaven National Laboratory, Upton, NY, USA*

[3]*Department of Physics, Faculty of Science, Al-Balqa Applied University, Salt, Jordan*

[4]*GSI Helmholtz Zentrum für Schwerionenforschung, Darmstadt, Germany*

[5]*Los Alamos National Laboratory, Los Alamos, USA*

[*]**reifarth@physik.uni-frankfurt.de**

**Abstract:**

The quest for the origin of the chemical elements, which we find in our body, in our planet (Earth), in our star (Sun), or in our galaxy (Milky Way) could only be resolved with a thorough understanding of the nuclear physics properties of stable and unstable atomic nuclei. While the elements until iron are either created during the big bang or during fusion reactions in stars, most of the elements heavier than iron are produced via neutron-induced reactions. Therefore, neutron capture cross sections of stable and unstable isotopes are important.

So far, time-of-flight or activation methods have been applied very successfully, but these methods reach their limits once the isotopes with half-lives shorter than a few months are of interest. A combination of a radioactive beam facility, an ion storage ring and a high flux reactor or a spallation source would allow a direct measurement of neutron-induced reactions over a wide energy range of isotopes with half-lives down to minutes.

The idea is to measure neutron-induced reactions on radioactive ions in inverse kinematics. This means, the radioactive ions will pass through a neutron target. In order to efficiently use the rare nuclides as well as to enhance the luminosity, the exotic nuclides can be stored in an ion storage ring. The neutron target can be the core of a research reactor, where one of the central fuel elements is replaced by the evacuated beam pipe of the storage ring. Alternatively, a large moderator surrounding a spallation source can be intersected by the beam pipe of an ion storage ring. Using particle detectors and Schottky spectroscopy, most of the important neutron-induced reactions, such as $(n,\gamma)$, $(n,p)$, $(n,\alpha)$, $(n,2n)$, or $(n,f)$, could be investigated.

1. **Introduction**

The quest for the origin of the chemical elements, which we find in our body, in our planet (Earth), in our star (Sun), or in our galaxy (Milky Way) could only be resolved with a thorough understanding of the nuclear physics properties of stable and unstable atomic nuclei [1, 2]. While the elements until iron are either created during the big bang or during fusion reactions in stars, most of the elements heavier than iron are produced via neutron-induced reactions, about equal contributions from the slow (s) and the rapid (r) neutron capture process [3], see Figure 1. Therefore, neutron capture cross sections of stable and unstable isotopes are important for a quantitative understanding.

During the s process, a given reaction can affect the abundance of only a few local isotopes, typically downstream of the reaction or a large number acting globally on the neutron exposure [4]. Neutron capture cross sections play an important role also during the r process. In hot r-process scenarios, they become important during the freeze-out phase [5, 6]. In cold r-process models like neutron star mergers neutron capture rates are important even during the merging process [7].

Even the production of the rare proton-rich isotopes can be affected by neutron capture processes. Those nuclei are typically produced by sequences of proton-induced or gamma-induced reactions starting from an s- or r-process seed [8]. Neutron captures on proton rich, short-lived nuclei affect the final abundance distribution of the important p nuclei $^{92,94}$Mo [9].

So far, time-of-flight or activation methods have been applied very successfully close or at the valley of stability, but these methods reach their limits once isotopes with half-lives shorter than a few months are of interest [10]. A combination of a radioactive beam facility, an ion storage ring and a high flux reactor or a spallation source would allow a direct measurement of neutron-induced reactions over a wide energy range of isotopes with half-lives down to minutes [11, 12].

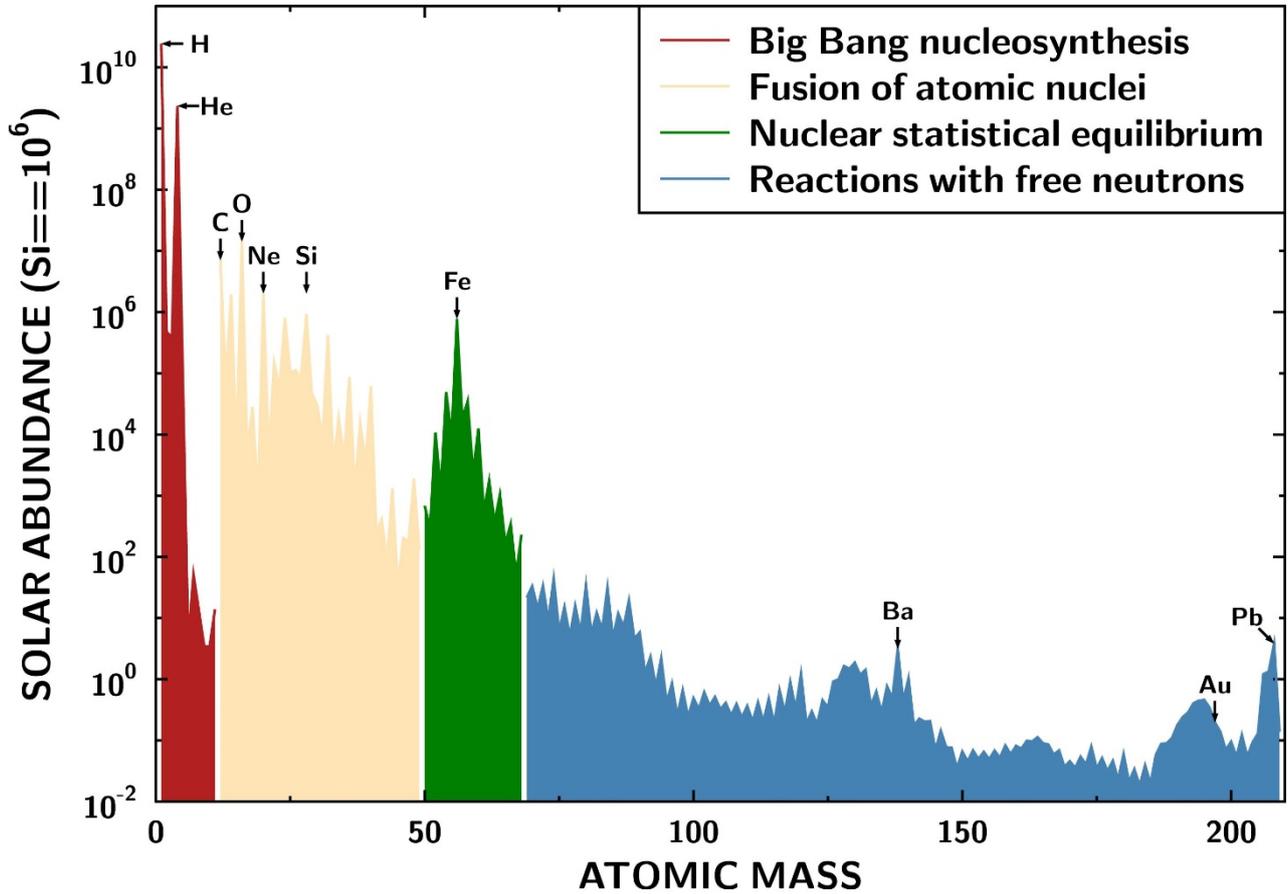

*Figure 1: The solar abundances of the atomic nuclei as a function of their mass [13]. The structures observed can be explained by different processes contributing to the overall nucleosynthesis. The lightest elements—hydrogen, helium—are produced in the Big Bang (red) while the elements up to iron are synthesized during stellar burning phases via fusion of charged particles (beige). During the extremely hot last stellar burning phase—the silicon burning—the isotopes around iron (mass 56), which are most tightly bound, are produced in the nuclear statistical equilibrium (green). Almost all of the elements with higher proton numbers than iron are produced through neutron-induced processes (blue).*

## 2. A spallation-based neutron target

The idea is to measure neutron-induced reactions on radioactive ions in inverse kinematics. This means, the radioactive ions will pass through a neutron gas. In contrast to charged-particle induced reaction rates like (p,γ) or (α,γ) where the inversion of kinematics implies a target of stable atoms like hydrogen or helium, the challenge is now to provide a target of free neutrons. In order to efficiently use the rare nuclides as well as to enhance the luminosity, the exotic nuclides should be stored in an ion storage ring [14]. The neutron target can be the core of a research reactor, where one of the central fuel elements is replaced by the evacuated beam pipe of the storage ring [11]. Alternatively, a large moderator surrounding a spallation source can be intersected by the beam pipe of an ion storage ring [12], see Figure 2. In this paper, we discuss parameter studies, which go beyond the basic concept published in the previous papers. In particular, the geometry of the moderator needs to be studied very carefully, since large amounts of valuable material are necessary. A second aspect is the design of the spallation target itself, which can be optimized in terms of geometry and material. All simulations presented here were performed with 4 different energies: 200 MeV, 800 MeV, 4 GeV, and 20 GeV. These energies were chosen, since they correspond to accelerators in operation or under construction, which might be used in combination with a spallation target. We will show later that the neutron density increases with the proton energy. Reduced beam energy can therefore be compensated by increased beam current.

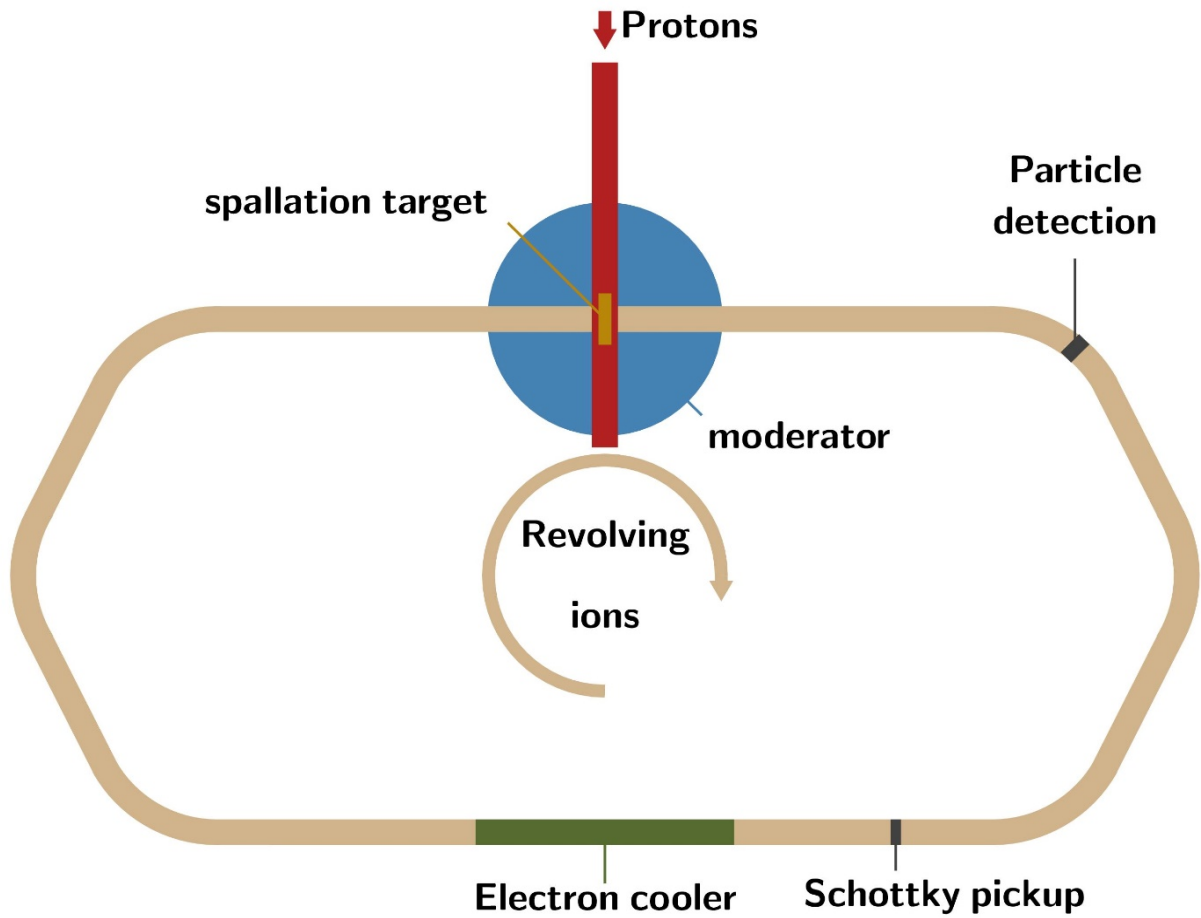

*Figure 2: Sketch of the proposed setup to investigate neutron-induced reactions in inverse kinematics. The isotopes under investigation will be stored in a ring, which penetrates a neutron moderator. This moderator acts as a trap for neutrons produced during a spallation reaction. Therefore, the neutron gas acts as a neutron target. See [12] for more details. The particle detectors can be silicon detectors just outside the unreacted beam [14].*

a. **Moderator material**

All the simulations published so far [12] have been performed assuming a heavy water moderator. However, other materials should be considered too, in particular if the moderator is not very large. Figure 3 shows the results for a 800 MeV proton beam hitting a small tungsten target. Different moderator materials and thicknesses have been investigated. It is very interesting to see that materials like beryllium or even regular (light) water have a better performance than heavy water for small moderators. But most materials reach a maximum neutron density already within the limits of the investigated moderator thicknesses. The explanation can be found in Figure 4. Already 1 m of moderator material absorbs a significant fraction of the neutrons, except for heavy water. Therefore, a further enlargement of the moderator does not improve the performance anymore since the neutrons are already absorbed inside the moderator, mostly via neutron captures on the moderator material. Beryllium is the best material below a moderator radius of 1.5 m. Deuterated polyethylene ($CD_2$) is the best option for moderators between 1.5 and 2 m radius. Heavy water is only the best option for very large moderators.

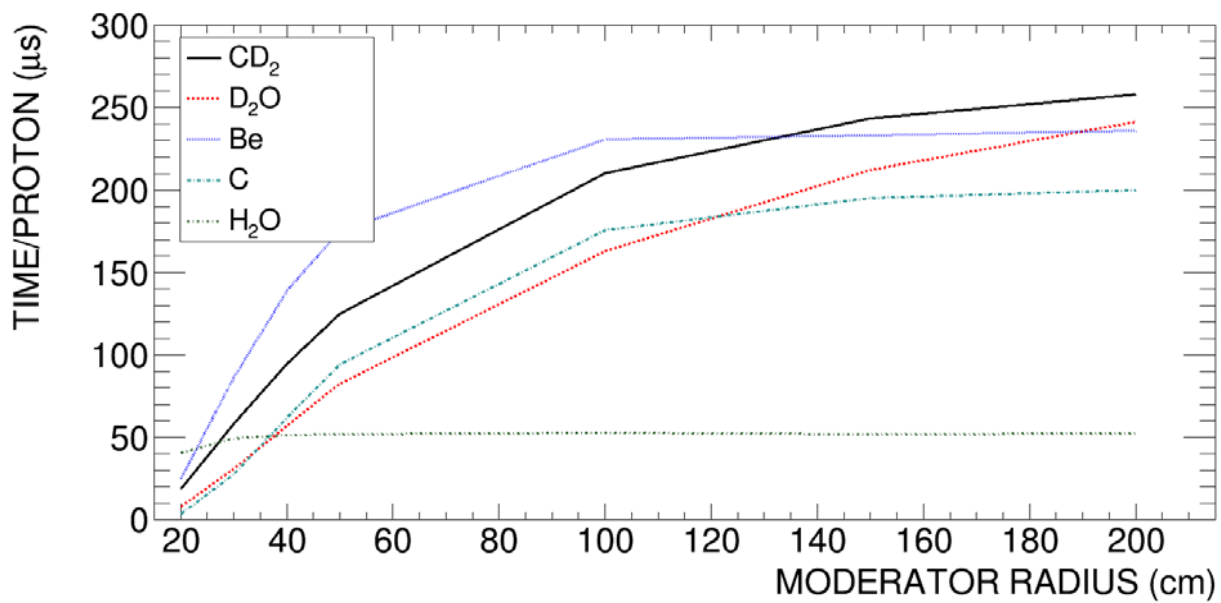

*Figure 3 : Total time per impinging proton spent by neutrons inside the beam pipe. The total time is directly proportional to the density of the neutron target. The simulations were performed with a proton beam of 800 MeV hitting a tungsten cylinder of 10 cm length and 3 cm diameter.*

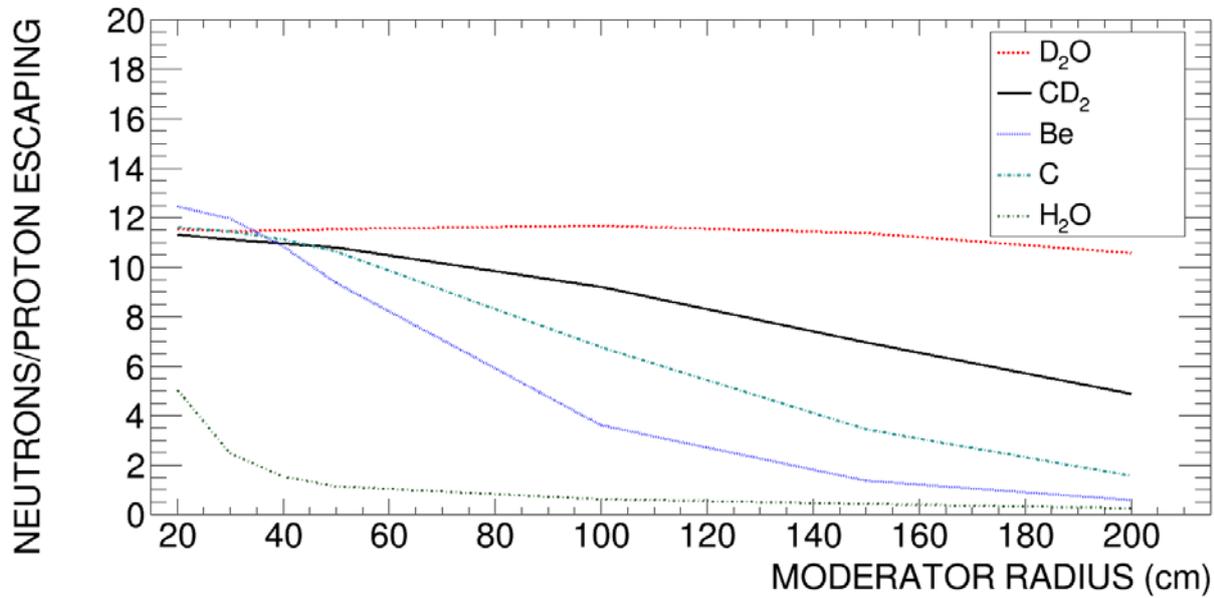

*Figure 4 : Number of neutrons escaping the moderator. The simulations were performed with a proton beam of 800 MeV hitting a tungsten cylinder of 10 cm length and 3 cm diameter.*

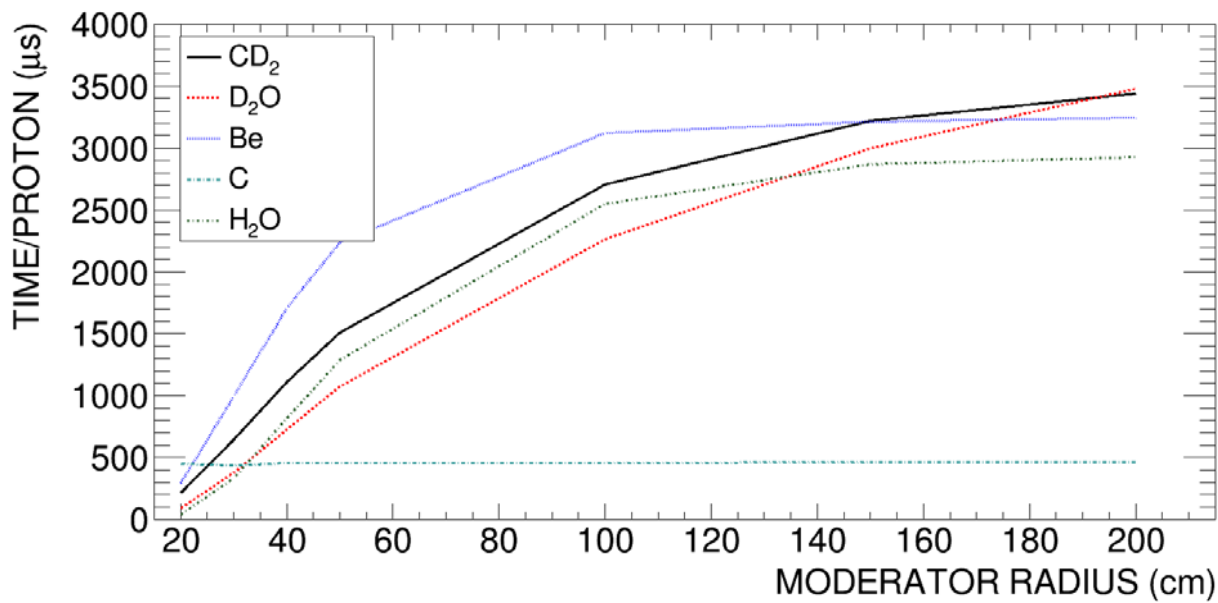

*Figure 5 : Total time per impinging proton spent by neutrons inside the beam pipe. The total time is directly proportional to the density of the neutron target. The simulations were performed with a proton beam of 20 GeV hitting a tungsten cylinder of 50 cm length and 4.8 cm diameter.*

A similar effect can be observed for a 20 GeV proton beam hitting a large tungsten target of 50 cm length and 4.8 cm diameter, Figure 5 and Figure 6. Also here, beryllium would be the best choice for moderators smaller than 1.5 m. Above 1.5 m $CD_2$ and $D_2O$ yield almost the same neutron density.

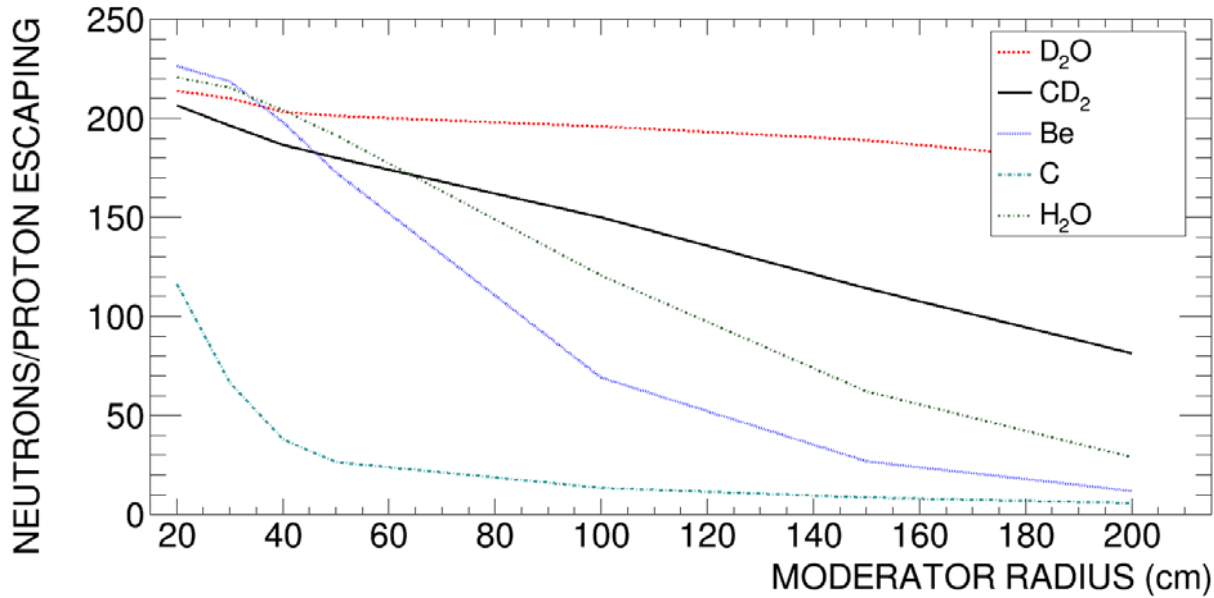

*Figure 6 : Number of neutrons escaping the moderator. The simulations were performed with a proton beam of 20 GeV hitting a tungsten cylinder of 50 cm length and 4.8 cm diameter.*

At lower proton energies of 200 MeV, the turn-over between Be and $CD_2$ occurs already at a moderator radius of 1 m (Figure 7). The reduction of the total scale is important to recognize. The maximum time spent by neutrons in the ion beam pipe is about 35 µs / proton in the case of a 200 MeV beam and close to 3500 µs / proton in the case of a 20 GeV beam. However, it is interesting to note that the maximum number of spallation neutrons per proton energy is reached around 1 GeV.

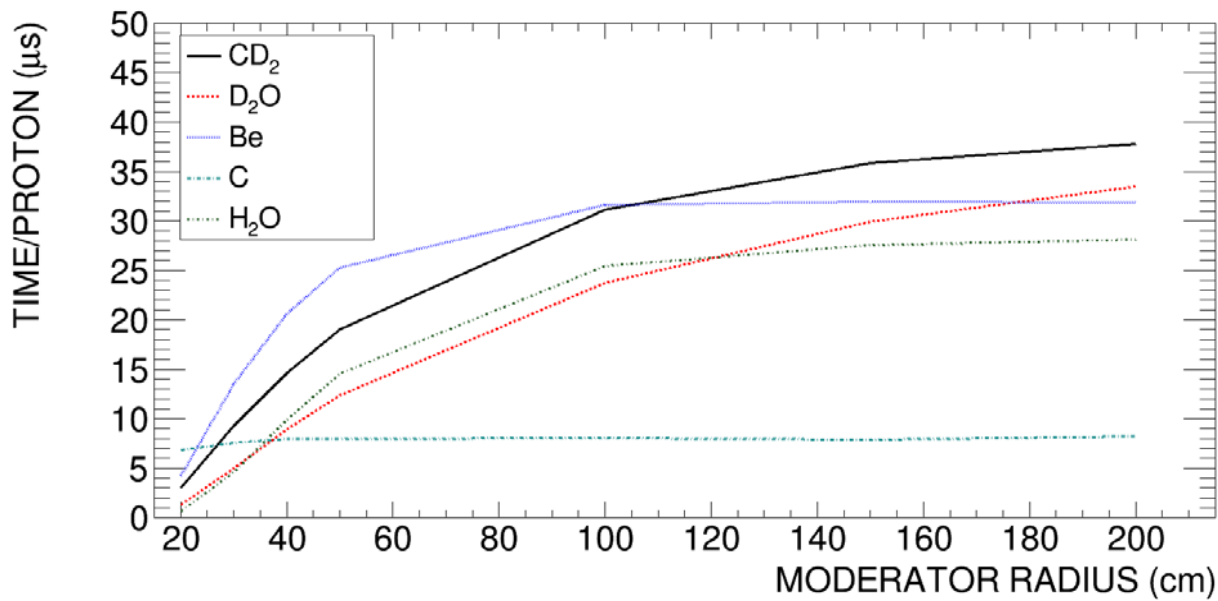

*Figure 7 : Total time per impinging proton spent by neutrons inside the beam pipe. The total time is directly proportional to the density of the neutron target. The simulations were performed with a proton beam of 200 MeV hitting a tungsten cylinder of 3 cm length and 3 cm diameter.*

We also investigated the idea of combining different moderator materials in shells. But it turned out that pure moderators always perform always better, see for example Figure 8.

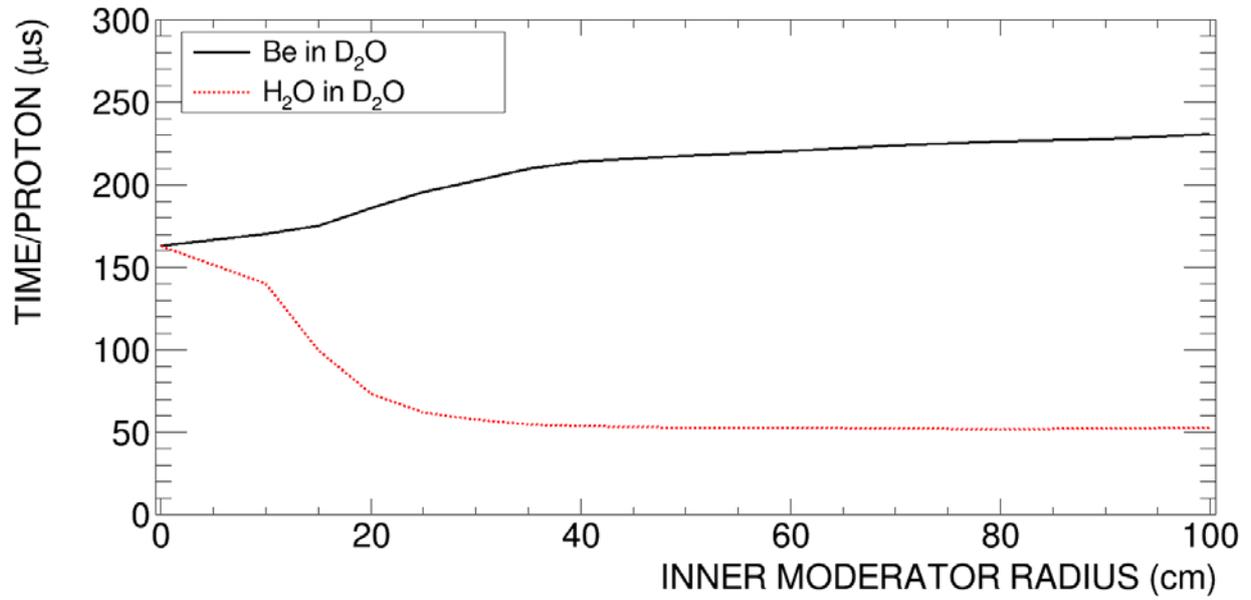

*Figure 8 : Investigation of a layered moderator. The outer radius was fixed at 1 m and the outer shell was heavy water. The radius of the inner shell varied between 0 and 1 m. Hence, the point a 0.0 corresponds to a pure heavy water moderator of 1 m. The points at 1 m correspond to pure moderators of Be and light water.*

b. **Spallation target**

The material as well as the geometry of the spallation target itself have an impact on the number of neutrons inside the ion beam pipe. Figure 9 shows a set of simulations, where the length of the spallation target was varied for two high-Z materials, tungsten and bismuth. The beam energies were 4 and 20 GeV. The overall performance of the two materials is very similar, but because of the lower density, the optimal target length for bismuth is larger than for tungsten. Figure 10 and Figure 11 show the results of similar investigations but for beam energies of 800 MeV and 200 MeV.

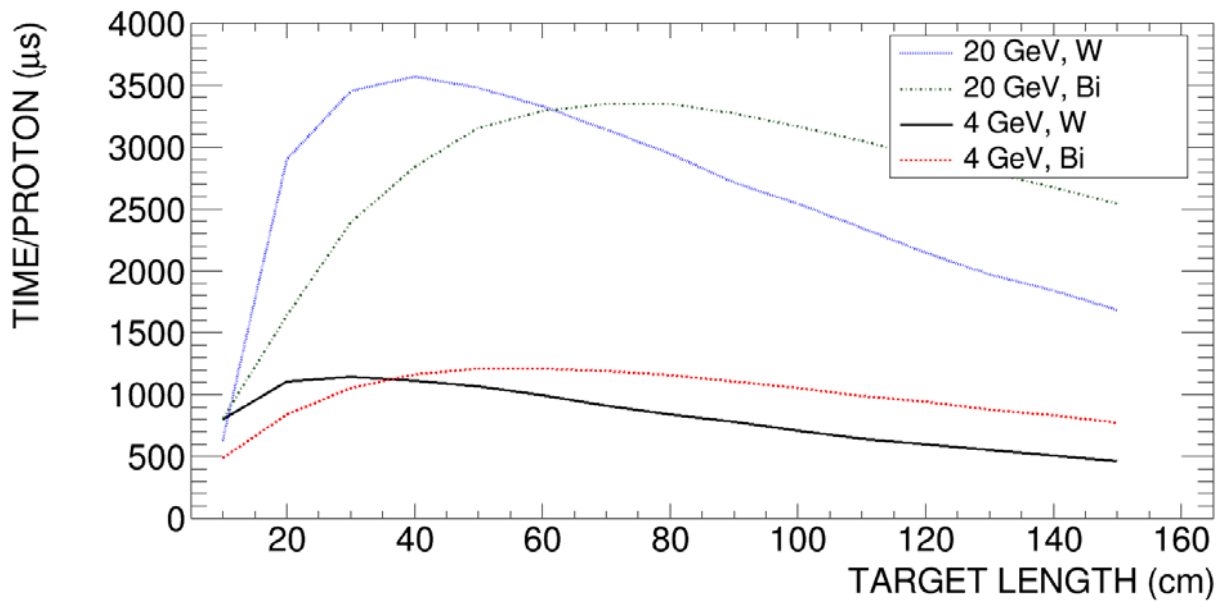

*Figure 9 : Total time per impinging proton spent by neutrons inside the beam pipe. The total time is directly proportional to the density of the neutron target. The simulations were performed with proton beams of 4 GeV and 20 GeV hitting a tungsten or bismuth cylinder of 4.8 cm diameter and different lengths. The moderator for this set of simulations was always $D_2O$ with a radius of 2 m.*

Reifarth et al, Neutron-induced reactions in nuclear astrophysics

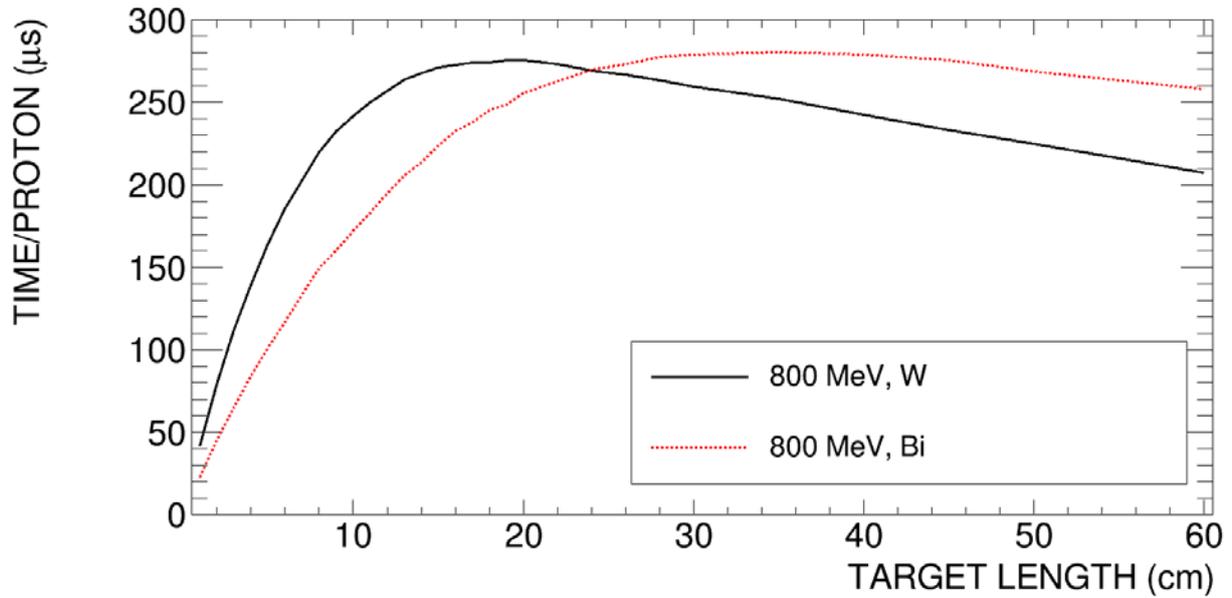

*Figure 10 : Total time per impinging proton spent by neutrons inside the beam pipe. The total time is directly proportional to the density of the neutron target. The simulations were performed with a proton beam of 800 MeV hitting a tungsten or bismuth cylinder of 3 cm diameter and different lengths. The moderator for this set of simulations was always D$_2$O with a radius of 2 m.*

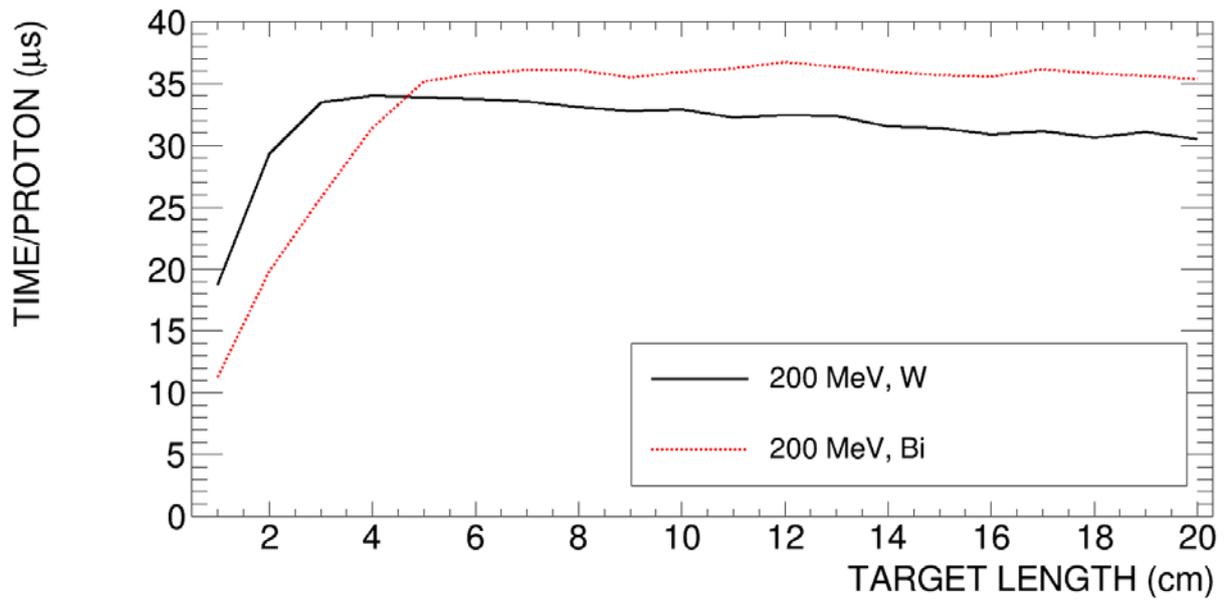

*Figure 11: Total time per impinging proton spent by neutrons inside the beam pipe. The total time is directly proportional to the density of the neutron target. The simulations were performed with a proton beam of 200 MeV hitting a tungsten or bismuth cylinder of 3 cm diameter and different lengths. The moderator for this set of simulations was always D$_2$O with a radius of 2 m.*

## 3. Conclusions

In order to estimate the actual density of the neutron target as seen by the revolving ions, one needs to multiply the times spent by neutrons as given in the figures above by the number of protons per time and divide by the cross section of the beam pipe. The simulated beam pipe has a diameter of 5 cm corresponding to a cross section of about 20 cm$^2$.

Table 1 shows the maximum values of the times/proton spent by neutrons inside the ion beam pipe for different beam energies, tungsten target and a 2 m moderator. In addition, the corresponding neutron densities per µA beam current are listed. Since the maximum of the neutron production per beam energy is around 1 GeV, the recommended setup would consist of an accelerator of about 1 GeV and a tungsten target of about 20 cm length.

*Table 1 : Number of neutrons/proton entering the moderator, maximum neutron time and density inside the ion beam pipe assuming a tungsten target, a $D_2O$ moderator of 2 m radius, and 1 µA beam current.*

| Proton energy (GeV) | Neutrons/Proton (produced) | Time per proton (µs) | Neutron density (cm$^{-2}$ µA$^{-1}$) |
|---|---|---|---|
| 0.2 | 1.2 | 34 | 1.1 10$^7$ |
| 0.8 | 12 | 275 | 8.8 10$^7$ |
| 4 | 54 | 1140 | 3.6 10$^8$ |
| 20 | 176 | 3570 | 1.1 10$^9$ |

The achievable vacuum conditions in the ring, the charge number and the charge state of the investigated ions determine the life time of the stored ions. Interactions with the rest gas will determine the lowest energies in the center of mass. A highly-optimized ring with reasonably low charge states should be able to store heavy ions down to 100 AkeV. This would cover the energy regime of hot, explosive stellar environments like the γ process or the r process. Even the second phase of the weak s process in massive stars occurring during the carbon shell burning temperatures around 90 keV are reached [15]. The restrictions on half-life of the isotopes is given by the requirement of a reasonably high duty cycle. The radioactive ions need to be produced, the beam stored, cooled and slowed down and the measurement performed. The beam preparations require currently several minutes [14], hence the half-life limit would be in this range. The reactions, which are possible to investigate depend on the detection method. Using particle detectors and Schottky spectroscopy, most of the important neutron-induced reactions, such as (n,γ), (n,p), (n,α), (n,2n), or at higher energies even (n,f), can be investigated [11].

The proposed setup would therefore significantly contribute to our understanding of the origin of the elements heavier than iron.

## 4. Acknowledgements

This research has received funding from the European Research Council under the European Union's Seventh Framework Programme (FP/2007-2013)/ERC Grant Agreement No. 615126.